\documentclass{pinchcr}   

\usepackage{MnSymbol,pifont}
\renewcommand{\d}{\mathrm{d}}
\newcommand{\D}{\mathrm{d}}
\newcommand{\bfp}{{\vec{p}}}
\newcommand{\bfP}{{\vec{P}}}
\newcommand{\bfe}{{\vec{e}}}
\newcommand{\bfr}{{\vec{r}}}
\newcommand{\bfv}{{\vec{v}}}
\newcommand{\bfvhat}{{\vec{\hat{v}}}}
\newcommand{\bfn}{{\vec{n}}}
\newcommand{\pdp}{{\vec{p}\cdot\delta\vec{p}}}

\title{Les Houches Summer School.\\
~\\
Active Matter \\
and\\
Non-Equilibrium Statistical Physics}

\begin{document}
\graphicspath{{./figures/}}

\maketitle


\maintext
\chapter{Collective Motion in Active Materials : Model Experiments}

\begin{center}
{\bf  Olivier Dauchot\\}
Gulliver Laboratory, UMR CNRS 7083, ESPCI Paris, PSL University,\\
10 rue Vauquelin, 75005 Paris, France
\end{center}


\section{Introduction}

Active systems transform energy from their environment into some form of work. The whole living world, but also virtually all human made machines, belong to this class of systems. Active matter, in turn, is composed of many such individual active units, essentially identical, which all individually perform some work and interact with each other.  At the macroscopic level, these intrinsically out of equilibrium materials are prone to develop new and interesting macroscopic physics~\shortcite{Vicsek:2012ty,Bechinger:2016cf,Fodor:2018en}. \\

Identifying the phases of a material, and their properties, given the knowledge of its elementary constituents, is the realm of statistical physics, which has been extremely successful in doing so for systems at equilibrium. As emphasized in J. Kurchan's chapter in the present series, this success is the result of the conjunction of two independent sets of properties, namely \lq\lq being at equilibrium\rq\rq , which is related to some form of time reversal symmetry, and \lq\lq being macroscopic\rq\rq, which often guarantees well behaved self-averaging properties.  In the case of active matter, the time reversal symmetry is broken at the microscopic level, and, for sure, principles of equilibrium statistical physics do not hold. On the contrary, there is no reason why macroscopically well defined states would not exist, hence the motivation for predicting them from minimal rules; a quest which has attracted the attention of a growing community of physicists, chemists and engineers.\\

To proceed further, we need to be more specific about the type of active material we have in mind. 
Looking for symmetries and conserved quantities (see. table~\ref{table:conservationlaws}) is a good way to classify the systems we are interested in. Standard materials are composed of molecules, the dynamics of which, prescribed by some Hamiltonian, conserves the number of particles, the energy and the momenta. It is clear that in active materials energy is dissipated at the level of each individual units and is therefore not conserved. In the following, we shall focus on the case where the work performed by the individual units ensures them self-propulsion, that is each unit gains or loses momentum, by exerting forces on its environment. This environment can be a solid substrate,  which directly serves as a source or sink of momentum, or a suspending fluid. In the later case, if the self-propelled particles and the fluid are away from any supporting walls, the momentum of the suspension including the fluid and the particles is conserved. Note that it is not the case if the fluid, in turn, exerts forces on a wall, which then takes the role of the substrate. In short, in the presence of a substrate the momentum is not conserved and one talks about \emph{dry} active matter wether there is a solvent or not; in the absence of substrate, the momentum of the set of self-propelled particles is again not conserved, but the one of the suspension, as a whole, is. One then talks about \emph{wet} active matter.\\

\begin{table}[t]
\center
\vspace{3mm}
\begin{tabular}{lccc}
\hline
 & Equilibrium & Wet Active Matter  & Dry Active Matter \\
\hline
Number of Particles  &  \ding{51} & \ding{51} & \ding{51}  \\
Energy 			& \ding{51} & \ding{55} & \ding{55}  \\
Momentum  		& \ding{51} & \ding{51} & \ding{55}  \\
\hline
\end{tabular}
\vspace{3mm}
\caption{Conserved (\ding{51}) and non-conserved (\ding{55}) quantities  in Dry vs Wet Active Matter as compared with Equilibrium systems. In this chapter we shall only consider dry active matter.}
\label{table:conservationlaws}
\end{table}

There are also symmetries associated with the self-propulsion of the particles. Consider a reference frame attached to the particle. The three main classes of motion are isotropic, directional and polar. In the isotropic case, the particle motion has no preferred direction; this is typically the case of granular media. In the polar case, the particle motion develops along one particular orientation in the reference frame attached to the particle, and in a particular direction along this orientation : the front/back symmetry is broken; this is the case of most animals. In the directional case, the motion has a preferred orientation, but the front/back symmetry is preserved.\\

In the present chapter, we shall concentrate on polar and dry active materials. The fluid phases of such materials exhibit two distinctive phenomenologies, which are specifically active, in the sense that none of them could take place at equilibrium.

First, when the interactions promote alignment of the velocities, a transition is observed from an isotropic dilute fluid, where particles move in all directions in a disordered way, to a long range ordered phase, where particles all move coherently in a given direction. This Transition to Collective Motion (TCM) has been first investigated in an effective model, the so-called Vicsek model~\shortcite{vicsek1995novel}. In this model, point particles move at constant speed and align their velocities when they encounter. The transition is controlled by the noise level and the density of particles. The transition is discontinuous and takes place via the nucleation of elongated propagating bands~\shortcite{Chate:2008isb,Solon:2013vr} (see also Hugues Chat\'e's chapter in the present series).
At large scale, the physics of the polar phase is described by an hydrodynamics theory either derived from conservation and symmetry arguments~\shortcite{toner1998flocks,toner1995long} (see also John Toner's chapter in the present series), or explicitly following kinetic theory~\shortcite{bertin2006boltzmann,Peshkov:2014un}.

Second, when the interactions lead to an effective decrease of the particle velocities with increasing local density, a condensation transition -- akin to the equilibrium gas-liquid transition -- takes place, although the interactions are purely repulsive. This is the so-called Motility Induced Phase Separation (MIPS)~\shortcite{Cates:2014dn}.\\

In the past two decades, experimentalists have motorized virtually all soft matter systems, from actin filaments~\shortcite{schaller467polar} and microtubules~\shortcite{Sanchez:2012gt} to grains~\shortcite{Deseigne:2010gc,Kumar:2014wr}, droplets~\shortcite{Thutupalli:2011bv,Izri:2014fv} and colloids~\shortcite{Palacci:2010hk,Bricard:2013jq,Buttinoni:2013de,Palacci:2013eu,Yan:2016ko}. A number of collective behaviors have been reported, among which the formation of clusters of particles, moving collectively or not, coarsening or not. Only in some cases~\shortcite{schaller467polar,Deseigne:2010gc,Bricard:2013jq} a clear transition to collective motion takes place.  
In all such experimental systems the interactions have many possible different origins, including steric repulsion, hydrodynamics, electrostatic and chemical coupling. In large group of animals, where collective motions are also observed, even more complex interactions, including social rules, may play an important role. Hence a central question of interest : for a given experimental system, with a well characterized microscopic dynamics, what is the expected large scale dynamics? \\

\noindent
Answering such a question is in general a hard task, for a number of different reasons:
\begin{itemize}
\item As already emphasized, there is no equilibrium-like first principle, such as free energy minimization, to guide our intuition.
\item Mapping the experimental parameters onto the parameters of the microscopic but effective models is often not possible.
\item In principle both MIPS and TCM can take place, eventually competing, depending on the experimental parameters. There is yet no unified theory, even at the effective level, to describe such a situation.
\item A number of experimental factors, which have been deliberately omitted in the effective models, could end up being relevant and modify the whole scenario.\\
\end{itemize}

The purpose of this chapter is to illustrate the above considerations using two well controlled experimental system of dry active matter, in which a transition to collective motion has been reported.  In both cases we shall describe the microscopic dynamics of their individual components as well as their large scale physics. This will allow us to discuss how, and to what extent, the latter can be inferred from the microscopic rules. The first experimental system has been introduced in~\shortcite{Bricard:2013jq}. It consists in a system of micron-size colloidal rollers, which interact through hydrodynamics and electrostatic interactions. For a given set of experimental parameters, the rolling speed remains constant, and the interactions only reorient the particles. We will see that it is then possible to obtain quantitative predictions for the observed large scale dynamics, using kinetic theory. We will discuss the role the long range hydrodynamics interaction could play. The second experimental system has been introduced in~\shortcite{Deseigne:2010gc}. It is composed of vibrated granular disks, which have been designed in such a way that they perform persistent random walk. At first sight this system looks simpler. The disks interact by essentially elastic collisions, which are not expected to induce steric alignment.  We shall see however that alignment does take place because of a natural coupling between the disk velocity and the disk orientation, which are in this case, two distinct degrees of freedom at the particle level. We shall then show explicitly how to compute the aligning properties of such a system and discuss how they compare to the effective alignment rule introduced in the Vicsek model. The chapter makes an extensive use of the works presented in~\shortcite{Bricard:2013jq}, -- especially the supplementary material --  as well as in~\shortcite{Weber:2013bj} and~\shortcite{Lam:2015bp,Lam:2015jr}. Here, we shall concentrate on the theoretical discussion, simply summarizing the main experimental observations. The reader is invited to refer to the above references for details about the experimental set up and procedures. 

\section{Colloidal Rollers}

The system consists in spherical colloidal particles, which self propel on solid surfaces and are sensitive to the orientation of their neighbors. The self-propulsion results from an electro-hydrodynamic phenomenon referred to as the Quincke rotation~\shortcite{Quincke}. As we shall see below in more details, when an electric field is applied to an insulating  sphere immersed in a conducting  fluid, the  charge distribution at the sphere surface is unstable  to infinitesimal fluctuations above a critical field amplitude $E>E_{\rm Q}$. This spontaneous symmetry breaking results in a net electrostatic torque on the sphere, which thus rotates at a constant speed around a random direction transverse to the electric field. 
When the sphere lies on a rigid surface, this spinning motion turns into self-propulsion. The electric and hydrodynamic interactions rule the alignment and repulsion between the rolling colloids.

\subsection{Experimental set up and major observations}

The authors~\shortcite{Bricard:2013jq} used PMMA beads of radius $a=2.5\,\mu\rm m$ diluted in an AOT ionic surfactant solution in hexadecane oil,  filling the gap between two conducting glass slides. Once the particles are sedimented, they apply an electric field $E$, and  indeed observed  the  high-speed rolling motion of the colloids ($v_0\simeq 10^2-10^3 a/s$).  The motion of the rollers is confined to race-track channels of width ${500\,\mu\rm m}<W<5\, \rm mm$, by adding a patterned insulating film at the surface of the upper transparent electrodes.\\

\begin{figure}[t!] 
\includegraphics[width=0.95\columnwidth,trim=0mm 0mm 0mm 0mm,clip]{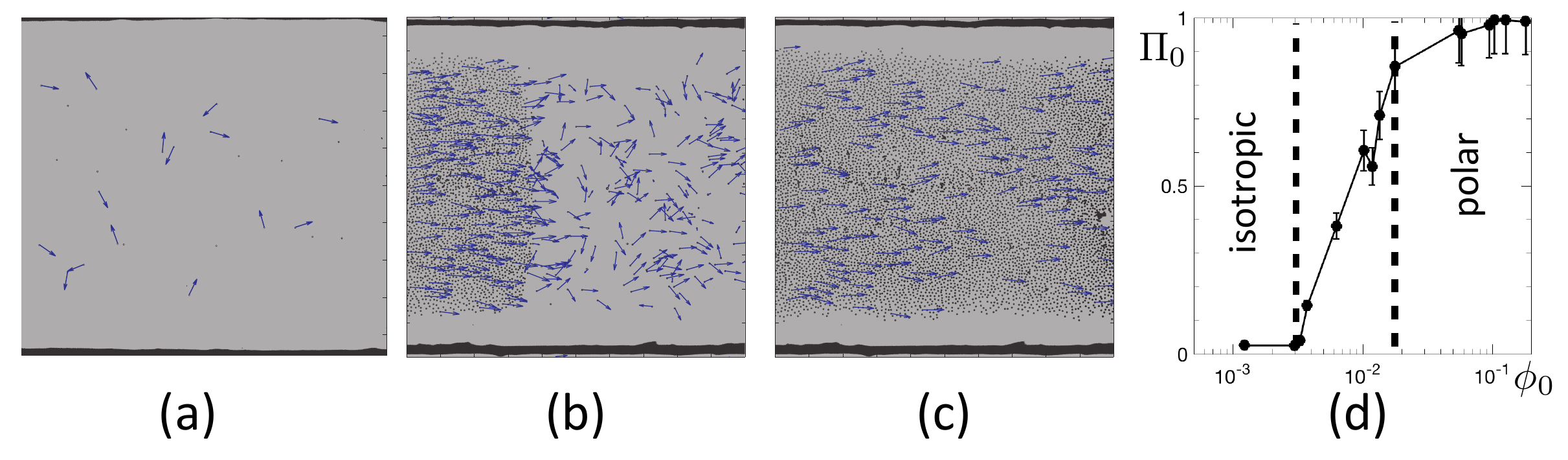}
\vspace{-0mm}
\caption{ {\bf Transition to collective motion in a system of rolling colloids, when increasing the area fraction $\phi_0$:} (a)~At low $\phi_0$ an isotropic gas is observed. (b)~Above a critical $\phi_c$, polar motion develops in the form of a macroscopic band (c)~At even larger $\phi_0$, one observes an homogeneous polar phase. (d)~Modulus of the average polarization $\Pi_0$, as a function of the area fraction $\phi_0$.}
\label{fig:rollers}
\vspace{-0mm}
\end{figure}

When their area fraction $\phi_0$ is small, the population of rollers behave as a gaseous phase (fig~\ref{fig:rollers}-a). All the particles move in random directions at the same speed. When the area fraction is increased above a critical value $\phi_{c}$, collective motion emerges spontaneously: a macroscopic fraction of the population moves in a coherent fashion along the same direction. 
For area fraction greater but close to  $\phi_c$, the system phase separates into an homogeneous isotropic phase and a denser polar phase, which typically consists in a single macroscopic band that propagates at a constant velocity $c_{\rm band}\simeq v_0$ (fig~\ref{fig:rollers}-b). The density within the band $\phi(s,t)$, where $s$ denotes the spatial coordinate along the track, sharply increases at the front and decays exponentially to a constant value which is very close to the critical volume fraction $\phi_c$ irrespective of the system size. The local polarization $\Pi(s,t)$, defined as the modulus of the averaged orientation of the velocities, is maximal at the front and decays continuously  to 0 along the band.  Quantitatively, it is observed that
$\phi(s,t)$ and $\Pi(s,t)$ are related by:
\begin{equation}
\label{eq:bands}
\Pi(s,t)=\left(1-\frac{\phi_c}{\phi(s,t)}\right)
\end{equation}
regardless of the particle velocity, and of the mean volume fraction. 
Further increasing the area fraction an homogeneous band-less polar state develops, with the the entire population of rollers cruising coherently along the same direction and $\Pi(s,t)\sim 1$ (fig~\ref{fig:rollers}-c).\\

These observations are very much reminiscent of the transition to collective motion reported in the Vicsek model~\shortcite{vicsek1995novel,Chate:2008isb,Solon:2013vr}. However, one specificity of the ordered polar phase obtained in the Vicseck model is the presence of anomalously large density fluctuations. Such fluctuations are also predicted at the level of the large-scale hydrodynamics theory~\shortcite{toner1998flocks,toner1995long}. Visual inspection of the polar phase of colloidal rollers does not provide any evidence of such density fluctuations. As a matter of fact, their first quantitative analysis concluded to their absence~\shortcite{Bricard:2013jq}; it is only recently, that it was shown that they are indeed present, conducting new experiments with larger statistics and weaker confinement~\shortcite{Geyer:2018in}.

\subsection{Microscopic dynamics : rolling mechanism and interactions}
The first step in analyzing theoretically this system, is to describe the self-propulsion mechanism and the interactions among the colloids. We shall recast here the main arguments of the derivation, which is provided in full details in the Supplementary Materials of~\shortcite{Bricard:2013jq}.\\

\subsubsection{Rolling mechanism}

\begin{figure}[t!] 
\includegraphics[width=0.95\columnwidth,trim=0mm 0mm 0mm 0mm,clip]{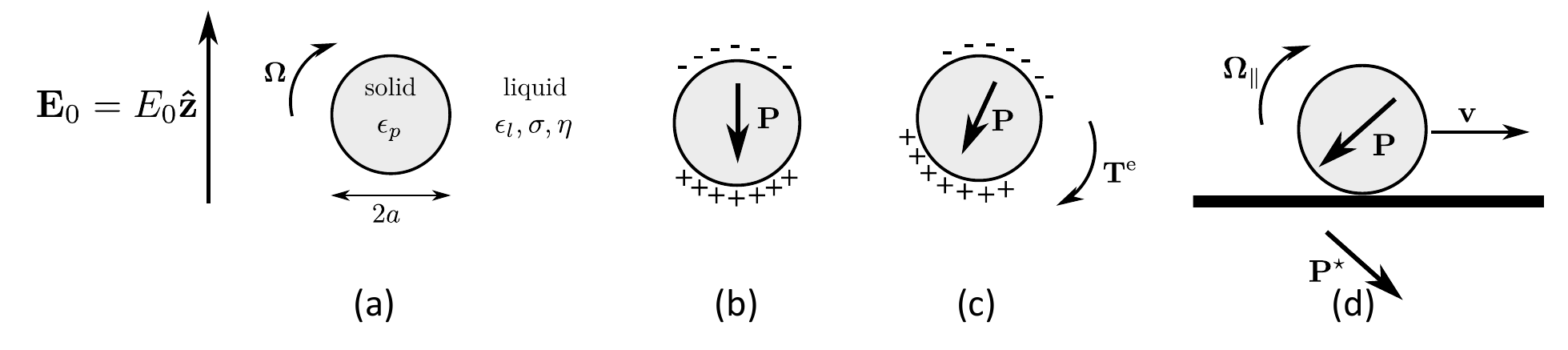}
\vspace{-0mm}
\caption{ {\bf Self propulsion mechanism:} (a)~An isolated solid and isolating sphere in an unbounded conducting liquid can undergo Quincke rotation, when an external electric field of amplitude $E_0>E_Q$ is applied. (b)~Electric charges accumulate at the particle-liquid interface and result in a dipolar surface distribution. (c)~A small rotational perturbation advects the charges located in the viscous layer and thereby tilts the dipole $\vec P$, thereby inducing a net electric torque $\vec T^{\rm e}$ which amplifies the initial perturbation. (d)~When laying on a iso-potential surface, the latter couples rotational and translational motion, allowing propulsion. It also disturbs the electric field. The dominant contribution to the image charge distribution is the symmetric dipole $\vec P^\star$.}
\label{fig:Quincke}
\vspace{-0mm}
\end{figure}

The Quincke rotation arises from the interplay between interfacial electrodynamics and the particle motion in a viscous fluid (see fig.~\ref{fig:Quincke}).  Let us  consider an insulating and impermeable sphere of radius $a$ located at $\vec r=\vec 0$. 
Here, for simplicity, we assume that the particle and the surrounding liquid share the same dielectric permittivity $\epsilon$ and the liquid has a conductivity $\sigma$.
An uniform DC electric field $E_0 \, \vec{\hat z}$ is applied along the $z$-direction.
Due to the conductivity discontinuity, a non-uniform charge distribution arises close to the interface. 
Assuming the thickness of the charge layer is much smaller than the particle radius $a$, it can be modeled by a  surface-charge distribution deduced from the continuity relation: $q_s = \epsilon(\vec E^l - \vec E^p) \cdot \vec{\hat r} \vert_{r=a}$, where $\vec E^l$ (resp. $\vec E^p$) stands for the electrostatic field inside the liquid (resp. the particle).
The surface charge conservation equation reads  $\partial_t \, q_s + \nabla_s \cdot \vec j_s = 0$, where $\nabla_s$  is the surface divergence operator, and $\vec j_s$ is the surface current. If the particle rotates with an angular velocity $\Omega$, both ohmic conduction and charge advection contribute to the surface current: $\vec j_s = \sigma \vec E + q_s \, \vec \Omega \times a \vec{\hat r}$.  The charge-conservation equation can finally be recast  into a dynamical equation for the first moment of the dipolar charge distribution $\vec P$:
\begin{equation}
\label{eqPtot}
        \frac{\d \vec P}{\d t} + \frac{1}{\tau} \vec P  = - \frac{1}{\tau}2 \pi \epsilon_0 a^3 \vec E_0 + \vec \Omega \times \vec P,
\end{equation}
where $\tau \equiv \frac{3 \epsilon}{2 \sigma}$ is the so-called Maxwell-Wagner time. When no rotation occurs, the dipole $\vec P$ relaxes towards a stationary value and orients along $- \vec E_0$ in a time $\tau$ due to the finite conductivity of the solution. However, as the particle rotates, charge advection competes with the spontaneous relaxation, and could in principle result in a dipole orientation making a finite angle with the external field $\vec E_0$.
%
The particle then experiences a net electric torque $\vec T^{\rm e} = \frac{\epsilon}{\epsilon_0} \vec P \times \vec E_0$ and a net force $\vec F^{\rm e} = \frac{\epsilon}{\epsilon_0} (\vec P \cdot \nabla) \vec E_0$, which vanishes in a uniform external field.
Neglecting the inertia of the sphere, the translation velocity $\vec v$ and the rotation speed $\vec \Omega$ are linearly related to $\vec F^{\rm e}$ and $\vec T^{\rm e}$ through a mobility matrix $\boldsymbol{\mathcal M}$:
\begin{equation}
\label{mobility_unbounded}
        \begin{pmatrix} \frac{1}{a} \vec v \\ \vec \Omega \end{pmatrix} \equiv \boldsymbol{\mathcal M} \cdot \begin{pmatrix}a \vec F^{\rm e} \\ \vec T^{\rm e} \end{pmatrix}\quad \text{with} \quad \boldsymbol{\mathcal M} = \begin{pmatrix} \mu_t \boldsymbol{I} & 0 \\ 0 & \mu_r \boldsymbol{I} \end{pmatrix},
\end{equation}
and $\mu_t = (6 \pi \eta a^3)^{-1}$ and $\mu_r = (8 \pi \eta a^3)^{-1}$, where $\eta$ is the liquid viscosity and $\boldsymbol{I}$ is the identity matrix. Note that $\boldsymbol{\mathcal M}$ is diagonal in an unbounded fluid only.
Eqs.~(\ref{eqPtot}) and~(\ref{mobility_unbounded}) together fully capture the particle dynamics. 
When the external field $\vec E_0$ exceeds the threshold value $E_{\rm Q} = \left[ 2 \pi \epsilon a^3 \mu_r \tau \right]^{-1/2}$, the trivial non-rotating solution is unstable against rotational perturbations and the dynamics bifurcates towards a solution describing a steady rotation at angular velocity
\begin{equation}
        \Omega = \frac{1}{\tau} \sqrt{\left(\frac{E_0}{E_{\rm Q}}\right)^2 -1}
\end{equation}
The rotation axis can be any direction perpendicular to $\vec E_0$ as the symmetry is spontaneously broken.
In short, any infinitesimal perturbation results in an electrostatic torque. When the external field exceeds a threshold value $E_{\rm Q}$, this torque is large enough to advect the charges despite the stabilizing mechanism provided by the finite conductivity of the solution. The advection amplifies the initial perturbation until the viscous torque balances the electric torque.  When the stationary state is reached, the  particle  steadily rotates at $\Omega$ around an axis perpendicular to $\vec E_0$, the direction of which is set by the initial perturbation only.\\

In an unbounded fluid and a uniform electric field, the particle experiences no net force and thus have no translational velocity. To achieve  propulsion of the spherical particle, the basic idea is to let it roll on a plane surface. The presence of such a surface modifies both the mechanical and the electrostatic equations introduced above. From the mechanical point of view, the presence of the surface is accounted for by a modified mobility matrix. Its main feature, as compared to the unbounded fluid case, is that it now contains off-diagonal terms, which couple rotational and translational motion: they are responsible for the rolling motion. The mobility coefficients are also modified and can be computed in the context of the lubrication hypothesis. The surface at $z=0$ does not only modify the hydrodynamics of the  fluid, but also disturbs the electric field. Indeed, the particle lies on the lower electrode, which is an equipotential surface. 
This is taken into account by considering an electric image charge distribution in the region $z < 0$, which is dominated by a dipole $\vec P^\star = P_z \, \vec{\hat z} - \vec P_\parallel$ at $z = -a$, as sketched in Fig.~\ref{fig:Quincke}(d). 
It follows from the symmetry of the real charges and of the image charges with respect to the equipotential plane, that 
the surface induces no tangential force $\vec F^{\rm e}_\parallel$, and no perpendicular torque $T^{\rm e}_z = 0$ on the sphere.  However, as opposed to the case of the classical Quincke setup, the particle experiences an external electric field which is {\em not} uniform. It includes here a correction $\delta \vec E^\star$ induced by the image charges. One can show that at leading order in $\vert \delta \vec E^\star \vert/E_0$, the relation between the polarization, the electric torque and the electric force is not affected by the  substrate. As a result
the  equations of motion of an isolated sphere lying on a planar electrode reads
\begin{equation}
\label{eqPv}
        \vec v = - \frac{\epsilon}{\epsilon_0} a \tilde \mu_t E_0 \, \vec P_\parallel,
\end{equation}
where $\tilde \mu_t$ is the modified translational mobility.
The particle steadily rolls on the electrode at a velocity $\vec v$, which points in the  direction opposite to the parallel component of the electric polarization. When $E_0 > E_{\rm Q}$,  the rolling speed $v_0\equiv|\vec v|$ is proportional to the in-plane component of $\vec P$, and is given by
\begin{equation}
        v_0 = \frac{a \tilde \mu_t}{\mu_r \tau} \sqrt{\left( \frac{E_0}{E_{\rm Q}} \right)^2 -1}
\end{equation}

It is important to note that, since the rolling motion results from a spontaneous symmetry breaking, there is no intrinsic orientation of the particle. This situation is different from that of Janus particles, for which an orientation is imprinted in the particle and thus is a degree of freedom correlated to, but different from the the velocity orientation. Note that the present situation is closer to the effective description chosen in the Vicsek model.

\subsubsection{Interactions}
The rollers are coupled by electrostatic and hydrodynamic interactions.  Their surface-charge distribution  induces a field disturbance $\delta \vec E (\vec r, t)$ which may alter the polarization, and the velocity, of the surrounding particles. Moreover, as it moves a roller induces a nontrivial fluid motion around it. Therefore, all the rollers are advected by a flow field $\vec u_\parallel(\vec r,t)$ resulting from the motion of their neighbors.\\
 
To derive these interactions, one first needs to compute the rollers dynamics in heterogeneous electric and hydrodynamic fields following the same spirit as what we did when introducing the bottom wall. One finds again a dynamical equation for 
$\vec P$, complemented by a mechanical equation that relates the velocity of the particle to the forces and torques acting on it. For a dilute system, one shows that despite the interactions, $P_z$, $P_\parallel$ and the norm of the velocity relax towards their unperturbed value over the timescale~$\tau$. In a dilute population of interacting rollers all the particles propel themselves at the same speed. However, as anticipated the orientations of the particles are now coupled, and evolve on much longer timescales $\sim \tau/\epsilon$, where $\epsilon$ quantifies the amplitude of the field perturbations. At leading order in $\epsilon$ this slow orientational dynamics  takes a rather simple form
\begin{equation}
\label{eq_theta}
        \frac{\d \theta}{\d t} = - \frac{\partial }{\partial \theta_i}\left[  -\mu_h \vec{\hat{v}}_i\cdot \vec u_\parallel + \mu_E \vec{\hat{v}}_i\cdot  \delta \vec E_\parallel  \right]
\end{equation}
where $\mu_h$ and $\mu_E$ are respectively the hydrodynamics and electric mobilities.  In short, the roller experiences
a torque induced by the flow field until its velocity aligns with $\vec u_\parallel$. Similarly, the second term accounts for the electrostatic coupling: it causes the roller velocity to align with $ - \delta \vec E_\parallel$.\\

\begin{figure}[t]
\begin{center}
\includegraphics[width=0.95\columnwidth,trim=0mm 0mm 0mm 0mm,clip]{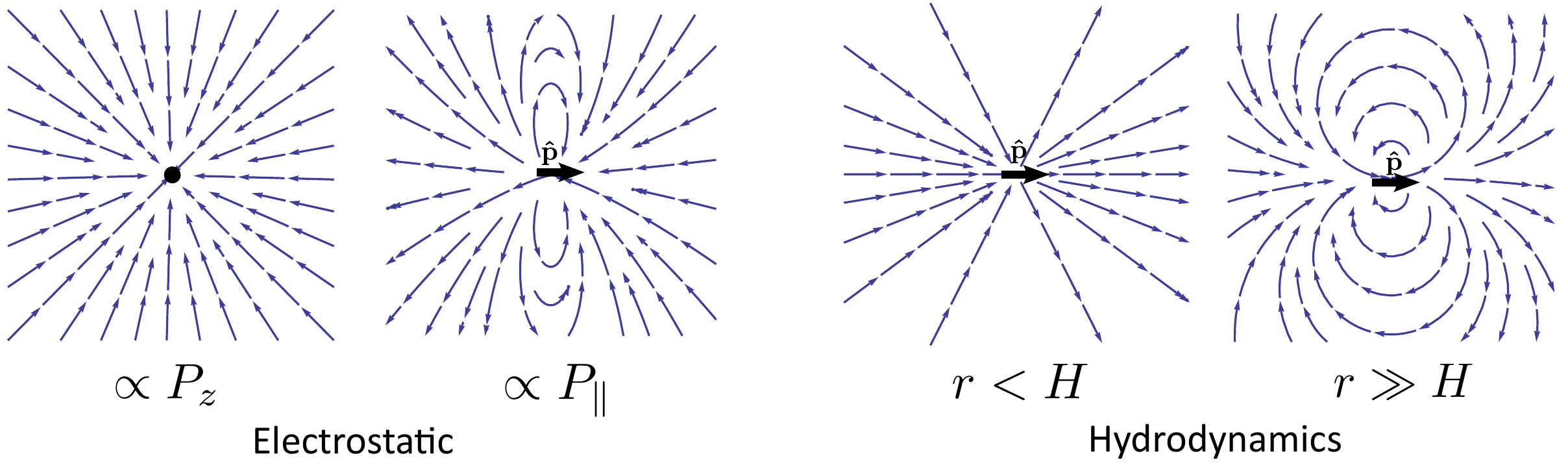} 
\end{center}
\caption{{\bf Electrostatic and Hydrodynamics interactions:} A particle rolling in direction $\vec{\hat p}$ creates a perturbative electric field. A radial part (proportional to $P_z$) results in a repulsive effect, which does not depend on the orientation of the  particle. An additional contribution (proportional to $P_\parallel$) breaks the rotational symmetry and yields a position-dependent interaction. The rolling particle also creates a perturbative flow field. At distances smaller than the channel height, the central roller induces a radial shear with anisotropic amplitude, which globally promotes alignment. At distances much larger than the channel height, the non-screened resulting flow has a dipolar symmetry.}
\label{fig:interactions}
\end{figure}

The next step is to derive explicitly the disturbance fields $\delta \vec E_\parallel(\vec r_i,t)$ and $\vec u_\parallel(\vec r_i,t)$ induced by all the other rollers $j \neq i$.  The electric field induced by a particle $j$ originates from the dipole $\vec P_j$ and its electrostatic image $\vec P^\star_j$. It has two contributions, which are displayed on figure~(\ref{fig:interactions})-left. A first contribution is proportional to $P_z$ and induces a repulsive interaction: it favors a roller velocity $\vec v_i$ pointing in the direction opposite to $\vec{\hat r}_{ij}$. The second contribution is proportional to $P_\parallel$, and it possibly results in alignment or anti-alignment with $\vec{\hat p}_j$, depending on the relative positions between the two rollers. 
At  distance larger than the separation distance $H$ between the two electrodes, all the electrostatic couplings are exponentially screened over a characteristic length $H/\pi$. Note that the electrostatic repulsion prevents the formation of dense regions of colloids. This justifies a posteriori the dilute limit treatment of the problem.
The flow field created by a rolling particle is expressed in terms of point-wise hydrodynamic singularities. Over distances smaller than the channel height $H$, a Quincke roller is akin to a rotlet near a no-slip wall: the particle is a point-wise torque-source. At long distances, unlike electrostatic screening, mass conservation gives rise to a non-vanishing flow having the form of a two-dimensional source dipole. The corresponding streamlines are plotted in Fig.~(\ref{fig:interactions})-right. 
At short distances, the hydrodynamic interactions promote the alignment of the roller velocities. In addition, long-range hydrodynamic interactions that algebraically decay as $r^{-2}$ have a dipolar symmetry. They can either cause alignment or anti-alignment, depending on the relative positions between the rollers.\\

\subsubsection{Equations of motion}
Assuming that both electrostatic and hydrodynamic interactions are pairwise additive, the above results can be summarized in a compact form. The particle $i$ moves at constant velocity $v_0$ on the surface, and undergoes a slow orientational dynamics:
\begin{align}
\label{eq_motion1}	&\vec{\dot r}_i = v_0 \vec{\hat p}_i\\
\label{eq_motion2}	&\dot \theta_i = \frac{1}{\tau} \frac{\partial}{\partial \theta_i} \sum_{j \neq i} \mathcal H_{\rm eff} (\vec r_i - \vec r_j, \vec{\hat p}_i,  \vec{\hat p}_j) 
\end{align}
The interaction potential $\mathcal H_{\rm eff}$ takes the form:
\begin{equation}
\label{potential}
        \mathcal H_{\rm eff} (\vec r, \vec{\hat p}_i,  \vec{\hat p}_j) = A(r) \, \vec{\hat p}_j \cdot \vec{\hat p}_i + B(r) \, \vec{\hat r} \cdot \vec{\hat p}_i + C(r) \, \vec{\hat p}_j \cdot (2\vec{\hat r}\vec{\hat r} - \vec I  )\cdot \vec{\hat p}_i
\end{equation}
where the coefficients have complex expressions, deduced from well identified microscopic parameters (see Supp. Mat in~\shortcite{Bricard:2013jq} for the exact expressions).\\

The term $A(r)$ describes the alignment interaction. It arises  both from the short-distance hydrodynamic interactions and from part of the electrostatic couplings.  The coefficient $B(r)$ is  positive, and corresponds to the electrostatic repulsive coupling. The last term $C(r)$ combines electric and hydrodynamic interactions. $A(r)$ and $B(r)$ are finite-range interactions, being screened on a distance set by the channel height. Conversely, $C(r)$ contains the unscreened dipolar hydrodynamic coupling. It is truly long-ranged since it algebraically decays like $r^{-2}$ in two dimensions. Note however, that its strength is small compared to the short-range hydrodynamic effect, since it is proportional to $a/H \ll 1$.\\

It is important to note that the interactions are far more complex than the alignment rules defined in the Vicsek model, for which, in the absence of noise, the incoming velocities fully align. The fact that more complex aligning interactions observed here lead to the same macroscopic dynamics is far from obvious. On one hand the symmetries of the microscopic dynamics are identical; on the other hand the short range repulsive and the long range dipolar interactions could in principle be relevant too, and impact the large scale dynamics.

\subsection{Deriving the large-scale hydrodynamics}
In order to discuss the relevance of the several interaction terms, one needs to derive the large scale dynamics from the microscopic rules using kinetic theory. We shall here only present the main steps of the procedure, which are as usual :
\begin{itemize}
\item Obtain the evolution equation for the $N$ particles probability density, from the dynamical equations;
\item Integrate out all but one particles, in order to derive the evolution equation for the $1$ particle probability density. This requires closing the associated hierarchy of equations.
\item Define the large scale fields, which are moments of this distribution. One should keep only slowly varying fields, that is those associated either to conserved quantities or to the slow modes associated with broken symmetries
\item Compute the evolution equation for these fields and close them with constitutive laws.
\end{itemize}

First, a noise term $\sqrt{2 D_r} \, \xi_i(t)$, where $\xi_i(t)$ is a Gaussian white noise with zero mean and unit variance, is added to Eq.~(\ref{eq_motion2}) in order to account for rotational diffusion. The resulting  $2N$ coupled Langevin equations~(\ref{eq_motion1})--(\ref{eq_motion2}) are then transformed into a Fokker-Planck equation for the $N$-particle distribution function $\Psi^{(N)}(\vec r_1..., \vec r_N, \theta_1, ..., \theta_N, t)$. 
By integrating over $N-1$ particle positions and directions, one obtains the time evolution of the one-particle density $\Psi^{(1)}(\vec r, \theta, t)$:
\begin{equation}
\label{Psi_1}
	\partial_t \Psi^{(1)} + v_0 \, \vec{\hat p} \cdot \nabla \Psi^{(1)} + \frac{1}{\tau} \partial_\theta  \int d^2 \vec r' \d \theta' \; \frac{\partial \mathcal H_{\rm eff} (\vec r - \vec r', \theta,  \theta')}{\partial \theta}  \Psi^{(2)}(\vec r, \vec r',  \theta, \theta', t) - D_r \, \partial_\theta^2 \Psi^{(1)} = 0,
\end{equation}
which depends on the two-point distribution function $\Psi^{(2)}(\vec r, \vec r',  \theta, \theta', t)$. 
This is the first equation of an infinite hierarchy, which couples the $n$-point  distribution $\Psi^{(n)}$ to the $(n+1)$-point distribution $\Psi^{(n+1)}$. This hierarchy of equations, must be closed, by postulating a relation between $\Psi^{(2)}$ and $\Psi^{(1)}$. Here we assume that the two-body correlations cancel over a distance as small as one particle diameter and include steric exclusion effects between the colloids:
\begin{equation}
\label{closure1}
	\Psi^{(2)}(\vec r, \vec r', \theta, \theta', t) = \begin{cases} 0 &\textrm{if } \vert \vec r - \vec r' \vert < 2a \\
	\Psi^{(1)}(\vec r, \theta, t) \Psi^{(1)}(\vec r', \theta', t) &\textrm{if } \vert \vec r - \vec r' \vert \geq 2a \end{cases}
\end{equation}
We thereby obtain from Eq.~(\ref{Psi_1}) a closed equation for the one-particle distribution function.\\

Second, one considers the three hydrodynamic fields:
\begin{alignat}{3}
	&\textrm{Area fraction:} \quad &&\phi(\vec r,t) \equiv \frac{1}{\pi a^2} \int \d \theta  \; \Psi^{(1)}(\vec r, \theta, t)\\
	&\textrm{Velocity polarization:} \quad&&\vec{\Pi}(\vec r,t) \equiv \frac{\pi a^2}{\phi} \int \d \theta  \; \vec{\hat p} \Psi^{(1)}(\vec r, \theta, t) \\
	&\textrm{Nematic order tensor:} \quad&&\vec Q(\vec r,t) \equiv \frac{\pi a^2}{\phi} \int \d \theta  \; \left( \vec{\hat p} \vec{\hat p} - \frac{1}{2}\vec I \right) \Psi^{(1)}(\vec r, \theta, t). 
\end{alignat}
By integrating Eq.~(\ref{Psi_1}), closed by Eq.~(\ref{closure1}), over $\theta$, one immediately recovers the particle-number conservation law:
\begin{equation}
\label{continuity}
	\partial_t \, \phi + v_0 \nabla \cdot (\phi \vec \Pi) = 0
\end{equation}
Taking the first angular moment of Eq.~(\ref{Psi_1}) similarly couples the time evolution of $\vec \Pi$ to the nematic order tensor~$\vec Q$. We thereby generate a new hierarchy of equations which couples each moment of the distribution function to higher-order moments.\\ 

The third and final steps thus consist in identifying one more closure assumption. It is important to understand that the closure assumption depends on the phase we want to describe. Indeed, it amounts to make some hypothesis on the shape of the angular distribution of the particles, which one expects to be very different in the isotropic phase and in the fully polarized phase. 

(i) Close to the instability threshold of the disordered state, one can use a Ginzburg-Landau-type expansion as introduced in~\shortcite{Bertin:2009bt} and further developed in \shortcite{Peshkov:2014un} (see also Hugues Chat\'e's chapter in the present series). Also assuming that the nematic order parameter $\vec Q$ relaxes faster than the two other hydrodynamics fields, one ends up with a close hydrodynamic equation for the velocity polarization $\Pi$ [Eq. (S29) of~\shortcite{Bricard:2013jq}]  coupled to the continuity equation~(\ref{continuity}). 

(ii) Far in the polar phase, it is necessary to introduce another closure approximation. In this highly polarized phase, one expects the angular probability distribution to be sharply peaked around the direction of collective motion. In this case, a simple assumption is to approximate it by the wrapped normal distribution. This "Gaussian" like Ansatz imposes the following relation $\vec Q = \Pi^2 \, \vec \Pi \, \vec \Pi - \frac{1}{2} \Pi^4 \, \vec I$. With this new closure relation, neglecting higher-order terms in $\frac{a}{H}$, one again obtains from Eqs.~(\ref{Psi_1})--(\ref{closure1}) an hydrodynamics equation for the velocity polarization $\Pi$ , [Eq. (S38) of~\shortcite{Bricard:2013jq}] coupled to the continuity equation~(\ref{continuity}). 

\subsection{Predictions and comparison to experiment}
We first discuss the transition to collective motion. Looking for homogeneous phases, one can drop space derivatives in the hydrodynamics equations. Close to threshold, the hydrodynamics equations then reduce to $\phi(\vec r,t)=\phi_0$ and
\begin{align}
\label{CW_isotropic}
	&\tau \partial_t \vec \Pi = \left(\alpha \, \phi_0 - \tau D_r \right) \vec \Pi - \frac{\alpha^2}{2 \tau D_r} (\phi_0^2 \Pi^2) \vec \Pi, 
\end{align}
where $\alpha \equiv \int_{r \geq 2a} \!\!\!\! \d r \; A(r) \frac{r}{a^2}$, accounts for the alignment interactions.
It readily follows from the cubic form of the r.h.s that the system undergoes a mean-field continuous phase transition to a polar state as $\phi_0$ exceeds the critical area fraction $\phi_c = \frac{\tau D_r}{\alpha}$. As the observed transition is not continuous, something else must take place. Indeed, one can perform the linear stability analysis of the homogeneous phases. As expected, the isotropic state is linearly stable for $\phi_0<\phi_c$ and unstable for $\phi_0>\phi_c$. What is more uncommon, is that the homogeneous weakly polarized state obtained for $\phi_0$ slightly larger than $\phi_c$ is also linearly unstable against compression fluctuations. Hence all homogeneous phases are linearly unstable at the onset of collective motion. This situation is exactly that of the Vicsek model. In such a situation, the system converges to a non linear solution, here the propagating bands. It is difficult to derive analytically the shape of band-density profiles. However, the particle-number conservation provides a relation between the local density and the local polarization field when density excitations  propagate steadily. Looking for propagative solutions of the form $\phi = \phi(x - c_{\textrm{band}}t)$, $\Pi = \Pi(x - c_{\textrm{band}}t) \vec{\hat x}$ and integrating Eq.~(\ref{continuity}) over the transverse direction leads to the relation
\begin{equation}
	\Pi(s) = \frac{c_{\textrm{band}}}{v_0} \left( 1 - \frac{\phi_\infty}{\phi(s)} \right)
\end{equation}
where the integration constant $\phi_\infty$ is the area fraction far away from the band. This is precisely the relation~(\ref{eq:bands}) observed experimentally.  Note that it does not depend on any closure scheme at the hydrodynamic level, because it derives from the continuity equation.\\

Turning to the polar liquid phase, we now investigate the linear stability of homogeneous polar phases, with respect to spatial fluctuations, for densities $\phi_0 \gg \phi_c$. The compression mode, which was unstable close to the transition is stabilized by the electrostatic repulsion at higher densities. Physically the electrostatic repulsion, impedes the formation of  highly concentrated regions. Fluctuations having the form of bend modes are exponentially amplified by the hydrodynamics long range interaction. However, as obviously expected transverse confinement eliminates this instability. Also remember that the amplitude of this interactions scales as $a/H\ll 1$. Finally, the fastest rate, which corresponds to a pure splay mode, is negative: splay fluctuations are stabilized by the long-range hydrodynamic interactions. This last results allows us to discuss the presence or absence of giant density fluctuations. Giant density fluctuations are a consequence of the splay mode instability. From the present result, we should thus conclude that they are suppressed, as indeed first reported in~\shortcite{Bricard:2013jq}. However the previous conclusion is valid in the limit of small wave vectors $qH \ll 1$. At  distances smaller than $H$, the long-range dipolar interactions, that govern the density fluctuations at large scales, are  subdominant. As a consequence, deviations from the above prediction are expected below a crossover length $\zeta$, which depends on the numerical values of the respective amplitude of the different interaction terms. This presumably explains why, indications of the giant density fluctuations could be captured in~\shortcite{Geyer:2018in}. 

\subsection{Discussion}

Altogether the experimental results together with their theoretical analysis establish that colloidal rollers self-assemble into a prototypical polar active fluid. In the present case one can conclude that the complexity of the microscopic interactions, as compared to their simplistic effective formulation in the Vicsek model, does not play a significant role. Establishing this statement however required a complete analysis of the microscopic interactions, at least in term of symmetries, an explicit derivation of the hydrodynamics equations, under a number of assumptions, and a linear stability analysis of the steady state solutions. In particular we saw that a careful study of the impact of the long range hydrodynamics interaction on the splay mode instability was necessary to decide about the presence of the giant density fluctuations.\\

The ability of polar active fluids to support sound modes, regardless of whether the dynamics of their microscopic units is overdamped, is one of the most remarkable theoretical predictions for active fluids with broken rotational symmetry~\shortcite{toner1995long,toner1998flocks,Toner:2005bj}. In their recent study~\shortcite{Geyer:2018in}, the authors have provided an experimental demonstration of this counterintuitive prediction, and establish a generic method to measure the material constants of active fluids from their sound spectrum.\\

It is truly remarkable to have obtained in a real experimental system such a clear realization of the physics described by the Vicsek model at the effective level. We personally believe that the two ingredients at the root of this observation are (i) that the speed of the rolling colloids is amazingly constant, wether the particles interact or not; (ii) that the particles interact without colliding. The dynamics thus reduces to a slow dynamics of the velocity orientation, which is precisely what the Vicsek model describes. 

\section{Granular Walkers}

The system consists in millimeter-sized disks with a built-in oriented axis. When vibrated the disks perform a persistent random walk. The collisions on the other hand are strictly isotropic. Although there is no obvious source of alignment, large-scale collective streams were reported in collections of approximately a thousand disks moving on a carefully vibrated plate~\shortcite{Deseigne:2010gc,Deseigne:2012kn}.
In order to bypass the inherent difficulties of the experimental setup (limited number of disks, limited size of the vibrated plate, limited range of control parameters), a model for the motion and collisions of the polar disks was proposed~\shortcite{Weber:2013bj}, which accounts quantitatively for the experimental properties at the single and pair interaction level and agrees well with observations at the collective level. The phase diagram of this model shares important similarities with that of the Vicsek model : for large noise and low density one observes a disordered gas, for low noise and large density an homogeneous polar phase sets in. At the transition, solitary polar bands propagate in a disordered surrounding phase.
The main question raised by this system is the origin of the microscopic alignment. Once it will be identified, we will discuss how it differs from the Vicsek one and the consequences, if any, at the macroscopic level.

\subsection{Experimental set up and major observations}
\begin{figure}[t]
\begin{center}
\includegraphics[width=0.95\columnwidth,trim=0mm 0mm 0mm 0mm,clip]{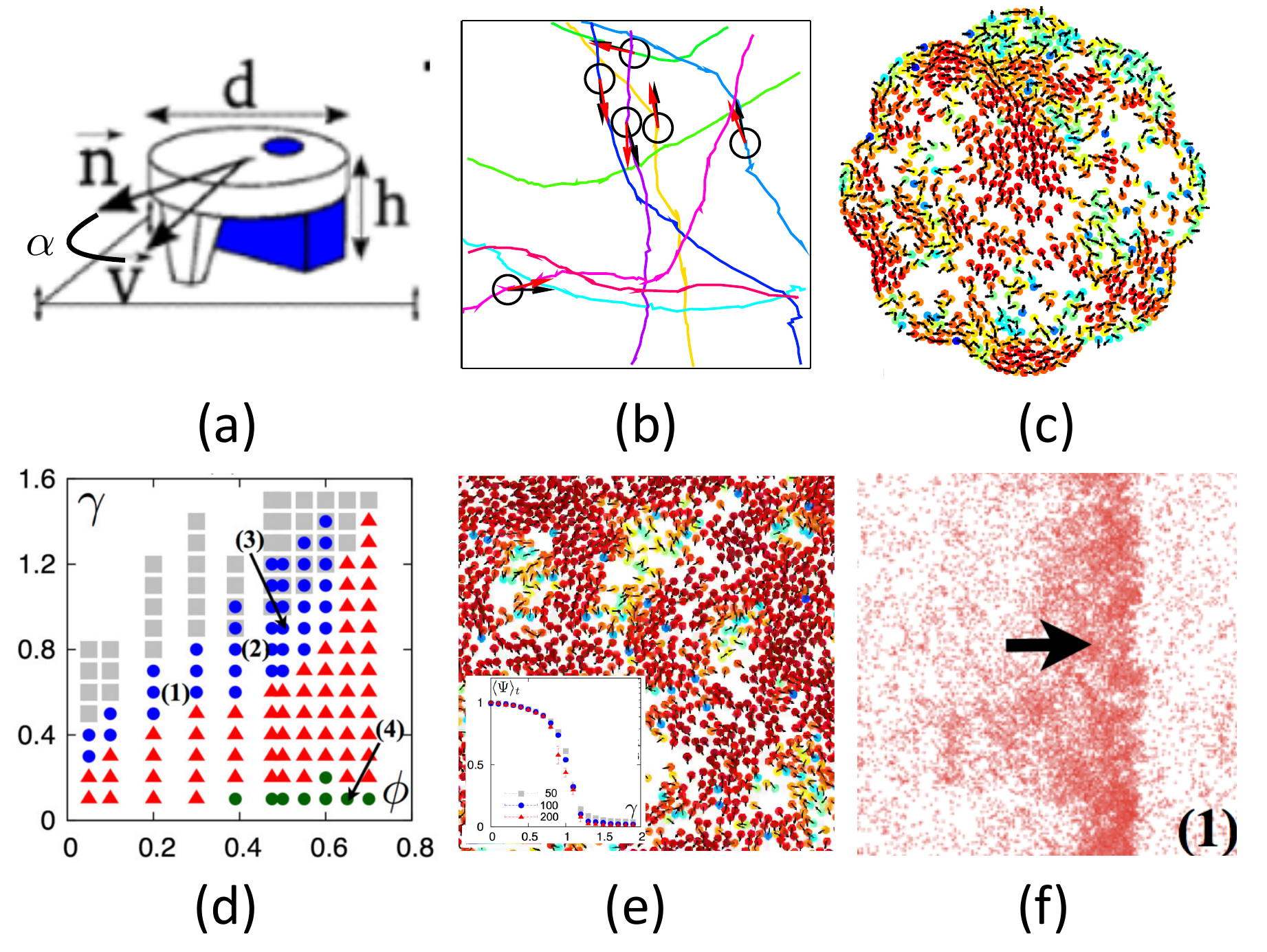} 
\end{center}
\caption{{\bf Transition to collective motion in a system of self-propelled walkers:} (a) Sketch of a walker : a hard metallic disc with an off-center tip and a glued rubber skate located at diametrically opposite positions; the velocity $\vec{v}$ is in general not perfectly aligned with the polarity $\vec n$. (b) Under proper vibration the walker performs a persistent random walk. (c) A thousand of such discs interacting through collisions develop large scale collective motions. (d) In silico mimicking the experimental system allows to explore the full phase diagram, with $\gamma$ the noise amplitude and $\phi$ the area fraction: (gray square = disordered phase; red triangles = homogeneous polar ordered phase; blue bullet = polar bands; green bullets = inverse polar bands). (e) Snapshot of the polar phase obtained with periodic boundary condition in a system of $N=1000$ particles (inset : order parameter as a function of $\gamma$; $\gamma=1$ corresponds to the experimental conditions). (f) Snapshot of a polar band state in a very large system ($L=200$). }
\label{fig:walkers}
\end{figure}

The polar particles are micro-machined copper-beryllium discs (diameter $d = 4$ mm) with an off-center tip and a glued rubber skate located at diametrically opposite positions.  These two "legs", with different mechanical response, endow the particles with a polar axis ($\vec n^i = (\cos\phi^i,\sin\phi^i)$). Of total height $h=2.0$ mm, the discs are sandwiched between two thick glass plates separated by a gap of $H=2.4$ mm (see Fig.\ref{fig:walkers}-a). 
Under proper vibration, the discs perform a persistent random walk, the persistence length of which is set by the vibration parameters (Fig.\ref{fig:walkers}-b).  Here we use a sinusoidal vibration of frequency $f=95$ Hz and relative acceleration to gravity $\Gamma = a (2\pi f)^2/g = 2.4$. Particle trajectories are tracked within a circular region of interest (ROI) of diameter $20d$, where the long-time averaged density field is homogeneous. For large enough vibration amplitude, $\Gamma>1$, individual velocities $\vec{v}_i(t) \equiv (\vec{r}_i(t+\tau_0)-\vec{r}_i(t))/\tau_0$ have a well-defined most probable or mean value $v_{\rm typ}\simeq 3 d/s$ , which depends only slightly on $\Gamma$. The local displacements of the particles are overwhelmingly taking place along $\vec{n}_i(t)$, their instantaneous polarity. The orientation angle diffuses, with an angular diffusion constant $D_\theta$, which increases fast and linearly  with $\Gamma$. As a result, the persistence length of an isolated polar particles $\xi\simeq 15 d$ for $\Gamma=2.4$ and decreases with increasing $\Gamma$.\\

Turning now to the collective dynamics of typically $N=1000$ particles and a packing fraction $\phi\simeq 0.40$, the authors~\shortcite{Deseigne:2010gc} report that at low $\Gamma$ values, for which the directed motion of the polar  particles is most persistent, they observe large-scale collective motion, with jets and swirls as large as the  system size. (Fig.\ref{fig:walkers}-c) Here because the boundary conditions are not periodic, the collective motion observed is not sustained at all times. Large moving clusters form, then breakdown, etc.  The orientational order is characterized by the modulus of the average velocity-defined polarity $\Psi(t) =| \langle \vec{u}_i(t) \rangle |$ where $\vec{u}_i(t)$ is the unit vector along  $\vec{v}_i(t)$ and  the average is over all particles inside the ROI at time $t$. The times series of the order parameter $\Psi$ presents strong variations, but can take a rather well-defined order one value for long periods of time.  At high $\Gamma$ values (large noise) no large-scale ordering is found. The study of the spatial and temporal correlation functions further confirm the onset of large scale collective motion.\\

In order to extend the observation range of the system, in terms of size and parameter values, a mathematical model for the motion and collisions of the polar disks was proposed~\shortcite{Weber:2013bj}.
Rather than modeling the full three-dimensional dynamics, the model describes the effective two-dimensional motion of the discs.  
As compared to the most common Active Brownian Particles (ABP), the main new features of the model, dictated by the experimental system, is twofold: 
(i) the dynamics of the particle's intrinsic polarity with respect to their velocity is explicitly described, and 
(ii) no explicit alignment rules are employed, but collisions are explicitly modeled.
Particle $i$ is subject to a noisy acceleration along its {\it polarity} axis $\vec{n}^i$  (with anisotropic, intrinsic, ``active" noise, respecting the particle's polar symmetry), balanced by an effective linear friction term along its {\it velocity} $\vec{v}^i=\frac{d}{dt} {\vec r}^i$, with $\vec{r}^i$ denoting the particle's coordinates.  Particles $i$ and $j$ with $|\vec r^i - \vec r^j|<d$, where $d$ is the particle diameter, interact by means of a pairwise, inelastic, repulsive interaction force $\vec F^{ij}_\epsilon$. Furthermore, when $\alpha^i= \widehat{(\vec{v}^i, \vec{n}^i)}$, the angle between velocity and polarity, is nonzero, frictional interactions with the vibrating plate are observed to induce a torque on the particle. More precisely, when $\alpha^i$, is acute, $\vec{n}^i$ rotates towards $\vec{v}^i$, while for $|\alpha^i|>\pi/2$, $\vec n^i$ rotates towards $-\vec{v}^i$. The model thus reads:
\begin{align}
\label{eq:dt_v}
& \frac{d}{dt}\vec{v}^{i} = [F_0+\eta_\parallel]  \vec{n}^i  + \eta_\perp \vec{n}^{i}_{\perp}
- \beta \vec{v}^i + \sum_{j}\vec F_\epsilon^{i,j},\\
\label{eq:dt_phi}
& \frac{d}{dt} \phi^i = \zeta \, \sin \alpha^i \, {\rm sign}(\cos \alpha^i) + \eta_\phi
\end{align}
where $F_0$, the propulsive force amplitude, and $\beta$, the damping coefficient, are constants giving rise to a stationary speed $v=F_0/\beta$. $\phi^i$ is the orientation of the unit vector $\vec{n}^i $ and $\zeta$ characterizes the strength of the coupling between polarity and velocity.
$\vec{n}^{i}_{\perp}$ is a unit vector perpendicular to $\vec n^i$, $\eta_{\parallel,\perp,\phi}$ represent Gaussian and wrapped Gaussian distributed white noises with zero mean, and $D_{\parallel,\perp,\phi}$ denotes the corresponding diffusion constant.\\

The numerical values of the coefficient of the model are first fixed by exploring the statistical properties of the one particle dynamics, namely its translational and angular diffusion properties, as well as the dynamical correlations between $\vec v$ and $\vec n$. Lastly the restitution coefficient $\epsilon$ is obtained by fitting the statistics of two-body collisions. The model provides a fair description of the experiment, also at the collective level.
Finally simulating the model with periodic boundary conditions, varying only the noise level and the packing fraction, a phase diagram akin to the Vicsek one was obtained, thereby establishing the first evidence of truly long range collective motion in an experimentalo-silico system of self-propelled particles Fig.\ref{fig:walkers}-d-e-f).\\

We however note two features distinct from the standard Vicsek case: (i) at large packing fraction and low noise, the authors report the existence of "inverse polar bands", where a dilute disordered region propagates within the polar phase; such "inverse polar bands" have never been reported in the Vicsek model; (ii) no polar bands could be observed when transitioning from the disordered to the polar phase, for packing fractions $\phi>0.6$.  The absence of polar bands in the transitional regime between the disordered state and the fully polar state might just be a finite-size effect; however, for the existing band, we note that the longitudinal density profile around $\phi \approx 0.6$ turns out to be rather flat, with an overall rather low order (as low as $\langle\psi\rangle_t\approx 0.2$ for $\phi=0.6$ and $\gamma=1.4$). These bands may thus be of different nature from the Vicsek-like, sharp, well-ordered bands found at low $\phi$, and could cease to exist asymptotically at a packing fraction below the rise of jamming and crystallization effects.

\subsection{Microscopic dynamics : reorientation mechanism}
In order to discuss the relation of the above observations with the Vicsek scenario, one first needs to understand the origin of the alignment: since the particles are discs, it is not expected to arise from steric interactions; it therefore must have a dynamical origin. In order to simplify the discussion let us first introduce the dimensionless and noiseless version of the above model:
\begin{align}
  \label{eq:dotbfr}
  {\dot \bfr}_i &= \bfv_i , \\
  \label{eq:dotbfv}
  \tau_v  {\dot \bfv}_i &= \bfn_i - \bfv_i + \sum_{j} \mathbf{f}_{ij}, \\
  \label{eq:dotbfn}
  \tau_n  {\dot \bfn}_i &= (\bfn_i \times \bfvhat_i) \times \bfn_i ,
\end{align}
where for simplicity, we consider that the torque in eq.~\ref{eq:dotbfn} always aligns $\bfv_i$ towards $\bfn_i$. Let us stress the presence of this torque is the key ingredient of the model and that it is a generic term for dry active systems. Even for a non chiral self-propulsion, namely when the propulsion mechanism is mirror-symmetric with respect to the body axis~$\bfn_i$, a torque acting on the particle is allowed by symmetries as soon as $\bfv_i$~is not aligned with~$\bfn_i$.
\begin{figure}[t]
\begin{center}
\includegraphics[width=0.95\columnwidth,trim=0mm 0mm 0mm 0mm,clip]{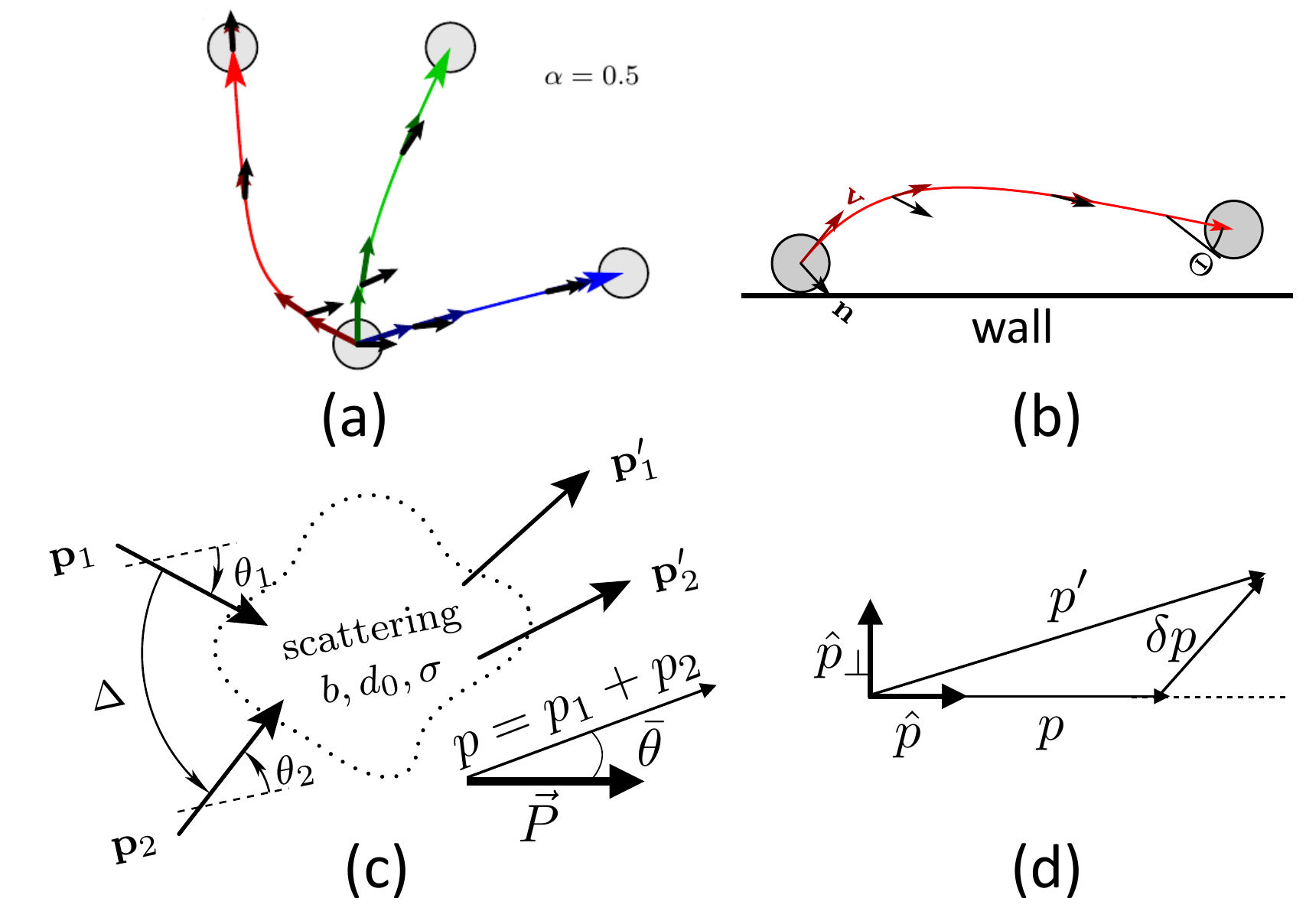} 
\end{center}
\caption{{\bf Dynamical alignment of self-propelled discs:} (a) When a particle starts with given $\bfv_i\neq\bfn_i$, the mutual relaxation towards each other causes the two vectors to converge against a common stationary direction, which mostly depends on $\alpha= \frac{\tau_n}{\tau_v}$, which can be understood as the persistence of the polarity vector~$\bfn$. (three trajectories with different initial conditions, for $\alpha=0.5$). (b) Trajectory of a particle following an elastic collision with a wall: for a finite $\alpha$, the mutual reorientation of $\bfv_i$ and $\bfn_i$ leads to an alignment of the trajectory with the wall. (c) Description of a scattering event. $\vec P$ is the total momentum in the system. Two particles incoming with momenta $p_1$ and $p_2$ exit the scattering event with momenta $p'_1$ and $p'_2$. The scattering event ends when all relaxation processes have taken place, namely the speeds have relaxed to $v_0$ and $\bfv_i\parallel\bfn_i$. (d) The alignment within the scattering event is positive if $pdp >0$.}
\label{fig:pdp}
\end{figure}
The mutual relaxation towards each other causes the two vectors to converge against a common stationary direction, where~$\bfv_i = v_0 \bfn_i$. Here, $v_0=F_0/\beta$ denotes the final speed of the isolated particle. The mass $m$ of the particles, their diameter~$d$, and~$d/v_0$, being respectively the units of mass, length and time, one has $\tau_v = \frac{m v_0}{\beta d}$ and $\tau_n=1/\zeta$. When a particle starts with given $\bfv_i\neq\bfn_i$, the trajectory depends on both parameters $\tau_v$ and~$\tau_n$. It turns out, however, that the final direction of the particle depends on these parameters mostly through their ratio $\alpha= \frac{\tau_n}{\tau_v}$ (not to be confused with the angle $\alpha$...), which can be understood as the persistence of the polarity vector~$\bfn$. 
When $\alpha\ll1$, then $\bfn$~aligns very fast and is practically always parallel to~$\bfv$. Conversely when $\alpha\gg1$, then $\bfn$~dictates the orientation of~$\bfv$. Quite remarkably the experiments conducted with the vibrated polar disks seem to work in the crossover between these limits, $\alpha\simeq 1$~\shortcite{Weber:2013bj}.

The dynamical origin of the alignment is summarized on figure~\ref{fig:pdp}-(a) and (b). Let's consider first the simpler case of the collision of a self-propelled disc with a wall. Before the collision, $\left| \bfv \right|=v_0$ and  $\bfv\parallel\bfn$. On the contrary, right after the collision,  the velocity has reoriented according to the collision rule, while $\bfn$ remains unchanged. During the mutual relaxation of $\bfv$ and $\bfn$, the trajectory is bended by the active force, pointing in the direction of $\bfn$, leading to an asymptotic common direction, which makes a smaller angle with the wall than the incoming direction, hence the alignment of the trajectory with the wall. Note that if $\alpha$ is large, the trajectory can also induce a new collision with the wall.  
Considering now the collision between two self-propelled discs, the situation is complicated by the fact that not all collisions are as simple as the one with the wall. In general they will depend on the impact parameter of the collision. The next step is therefore to consider all possible scattering events and somehow add their contribution to the global alignment.\\

Before doing so, let us mention, that on top of the deterministic equations~(\ref{eq:dotbfr}--\ref{eq:dotbfn}), one can add some angular noise distributed normally with zero mean and an angular diffusion constant $D$.

\subsection{Alignment rooted in the non conservation of momentum}
In this section, we shall derive a general setting to compute the average alignment for a system of self-propelled particles interacting by pairs. A published version of this discussion can be found in~\shortcite{Lam:2015jr} and further details in the supp. mat. of the arXiv https://lanl.arxiv.org/pdf/1410.4520v2.\\

Particle velocities at equilibrium obey the Maxwell-Boltzmann distribution; self-propelled particles do not. After some transient, a self-propelled particle reaches its intrinsic steady velocity $v_0$, set by the competition between propelling and dissipation mechanisms.  In the low-density limit, this transient lasts much less than the mean free flight time, and one can safely assume that particles have a constant speed $v_0$.  For spatially homogeneous states, the one-particle distribution thus reduces to the density probability $f(\theta,t)$ of having a particle with velocity $v_0\hat\bfe(\theta)$ at time~$t$, where $\hat\bfe(\theta)$ is the unit vector of polar angle~$\theta$.
This distribution evolves according to self-diffusion events and binary scattering events. 
A scattering event, as pictured schematically on Fig.~\ref{fig:pdp}-(c), is specified by the incoming angles $\theta_1$ and $\theta_2$ of the two particles or, equivalently, by the incoming half-angle \hbox{$\bar\theta = \mathrm{Arg}(e^{i\theta_1}{+}e^{i\theta_2})$} and the incoming angular separation \hbox{$\Delta = \theta_1{-}\theta_2$}.
Additional scattering {parameters}, such as the impact parameter, or some collisional noise, may be needed and are collectively noted as~$\zeta$.  A scattering event changes the momentum sum of the involved two particles by an amount~$\delta\bfp$, which depends {a priori} on all scattering parameters $\bar\theta$, $\Delta$ and $\zeta$.
The average momentum of all~$N$ particles in the system changes in this event from $\bfP$~into $\bfP'$, concluding that $N(\bfP'-\bfP)=\delta\bfp$.
In the same way, a self-diffusion event changes the momentum of a particle at $\theta_1$ by an amount $N(\bfP'-\bfP)=\delta\bfp_\mathrm{diff}(\theta_1,\eta) = \mathbf{R}_\eta \bfp - \bfp$, where $\mathbf{R}_\eta \bfp$ is the rotation of $\bfp=\hat\bfe(\theta_1)$ by an angle $\eta$.
The self-diffusion process is characterized by the probability density~$P_\eta(\eta)$ for a particle with angle $\theta_1$ to jump to angle $\theta_1+\eta$.
Assuming molecular chaos and averaging these two balance equations over the statistics of scattering and self-diffusion events taking place in a small time interval, one obtains the evolution equation by taking the continuous time limit:
\begin{align}
\label{eq:kin1}
\frac{\D\bfP}{\D t} &=  \lambda_\mathrm{diff}\: \Phi^{\mathrm{diff}}_f\Bigl[\delta\bfp_\mathrm{diff}(\theta_1,\eta)\Bigr] + \; \lambda\: \Phi^{\mathrm{scat}}_f\Bigl[\delta\bfp(\bar\theta,\Delta,\zeta)\Bigr] ,
\end{align}
where
\begin{align}
\label{eq:kin3}
\Phi^{\mathrm{diff}}_f\Bigl[\dots\Bigr]
&= \int_0^{2\pi} \D\theta_1 \int \D\eta \, P_\eta(\eta) \, f(\theta_1,t)\, (\dots),\\
\label{eq:kin2}
\Phi^{\mathrm{scat}}_f\Bigl[\dots\Bigr]
&=  \int_0^{2\pi}  \D\bar\theta \int_{-\pi}^{\pi} \D\Delta \int \D\zeta\, K(\Delta,\zeta) \, f(\theta_1,t)f(\theta_2,t)\, (\dots).
\end{align}
In the right hand side of Eq.~\eqref{eq:kin1}, the first term comes from the self-diffusion process, which happens at a characteristic rate $\lambda_{\mathrm{diff}}$.
The second term comes from the binary scattering process. In its integrand, a scattering event with scattering parameters $\theta_1$, $\theta_2$ and $\zeta$ is assumed to happen at a rate proportional to both $f(\theta_1,t)$ and $f(\theta_2,t)$; this comes from the molecular chaos hypothesis.
The proportionality factor is $\lambda K(\Delta,\zeta)$, the scattering rate of such an event. Note that it does not depend on $\bar\theta$ as a result of global rotational invariance.
As a convention, we have chosen to normalize $K$ such that $\frac{1}{2\pi}\int_{-\pi}^{\pi} \D\Delta \int\D\zeta K(\Delta,\zeta) = 1$. The prefactor $\lambda$ thus gives the characteristic scale of the scattering rate.
If one considers interacting disks with diameter $d_0$ at a number density $\rho$, a scattering event is entirely described by $\theta_1$, $\theta_2$ and the impact parameter $b$ (thus, $\int\D\zeta\equiv \int_{-d_0}^{d_0}\D b$). By using the construction of the Boltzmann cylinder, one finds for the scattering rate $\lambda K(\Delta,b) = \rho v_0 |\sin\frac\Delta2|$. Importantly, it is proportional to the density and does not depend on the impact parameter.\\

Equation~\eqref{eq:kin1} gives the evolution of the vectorial order parameter $\bfP$. Now, in order to get the evolution of $\psi=|\bfP|$, we go to polar coordinates $\bfP = \psi\,\hat\bfe(\theta_P)$ and project Eq.~\eqref{eq:kin1} on the radial direction~$\hat\bfe(\theta_P)$. In the absence of chirality, $\bfP$ keeps its angular direction, so that one can set $\theta_P(t)=0$.
As for the binary scattering term, we find for the projection $\Phi^{\mathrm{scat}}_f \bigl[\delta\bfp\bigr] \cdot \hat\bfe(\theta_P) = \Phi^{\mathrm{scat}}_f\bigl[(\hat\bfp \cdot \delta\bfp)\cos\bar\theta\bigr]$.
For the self-diffusion term, we can compute the integral explicitly and obtain $\lambda_\mathrm{diff}\Phi^{\mathrm{diff}}_f\bigl[\delta\bfp_\mathrm{diff}\bigr] = -D\psi$, where the self-diffusion constant is given by
\begin{equation}
D=\lambda_\mathrm{diff}\left(1-\int\D\eta\,P_\eta(\eta)\cos\eta\right) \ge 0 .
\end{equation}
Altogether, the radial component of Eq.~\eqref{eq:kin1} reads:
\begin{equation}
    \label{eq:dpsidt}
    \frac{\D\psi}{\D t} = \lambda \Phi^\mathrm{scat}_f\!\Big[(\hat\bfp \cdot\delta\bfp)\cos\bar\theta \Big] - D\psi .
\end{equation}
This evolution equation is derived from Eq.~\eqref{eq:kin1} with the only additional assumption that the system is not chiral. We keep this assumption in what follows.\\

As usual, the kinetic equation \eqref{eq:dpsidt} is of no use if the angular distribution $f(\theta,t)$ is unknown. The strategy consists in introducing an Ansatz for the distribution, assuming that the time-dependence is implicitly given by $\psi(t)$:
\begin{equation}
    f(\theta,t) = \Upsilon_{\psi(t)}(\theta)
\end{equation}
where $\Upsilon_\psi(\theta)$ for all $0\le\psi\le1$ is a family of angular distribution functions.
Without further specifying the shape of the distribution family $\Upsilon_{\psi}$, one readily obtains a close equation for the order parameter $\psi(t)$
\begin{equation}
\label{eq:dpsidt_ansatz_tot}
    \frac{\D\psi}{\D t} = \lambda F(\psi) - D \psi,
\end{equation}
with
\begin{equation}
\label{eq:dpsidt_ansatz}
    F(\psi)  =  \Phi_\psi[\hat\bfp\cdot\delta\bfp \cos\bar\theta] =
    \int_{-\pi}^{\pi} \frac{\D\Delta}{2\pi} \int \D \zeta \: K(\Delta,\zeta) \, g(\psi, \Delta) \, \pdp.
\end{equation}
and
\begin{equation}
\label{eq:gdef}
g(\psi,\Delta) =  \frac{2\pi}{2 \cos \frac{\Delta}{2} } \int_{0}^{2\pi} \D  \bar\theta
\, \Upsilon_\psi(\bar\theta+\Delta/2) \Upsilon_\psi(\bar\theta-\Delta/2) \cos\bar\theta
\end{equation}
Finally, we now specialize to the wrapped Gaussian or von Mises angular distribution, which is a generalization of the Gaussian on a periodic interval and the simplest distribution to describe fluctuations around a given orientation; it is given by
\begin{equation}    
\label{eq:vonmises}
	\Upsilon_\psi(\theta) = \frac{e^{\kappa(\psi) \cos\theta}}{2\pi I_0(\kappa(\psi))},
\end{equation}
where 
\begin{equation}
    I_n(x) = \frac1{2\pi} \int_0^{2\pi} \D \theta\, e^{x\cos\theta} \cos{n\theta}
\end{equation}
are the order $n$ modified Bessel function of the first kind and $K(\psi)$ is obtained from the inversion of the implicit relation  $\frac{I_1(\kappa)}{I_0(\kappa)} = \psi$.
The integrations over $\D\bar\theta$ in Eq.~\eqref{eq:gdef} can be performed explicitly to obtain
\begin{equation}
\label{eq:g:vonmises}
    g(\psi,\Delta) =
    \frac{\kappa(\psi)}{I_0^2(\kappa(\psi))}
    \frac{I_1\big(2\kappa(\psi)\cos\frac{\Delta}2\big)}{2\kappa(\psi)\cos\frac{\Delta}2}.
\end{equation}
Eqs.~(\ref{eq:dpsidt_ansatz_tot}),~(\ref{eq:dpsidt_ansatz}) and~(\ref{eq:g:vonmises}) provide an explicit solution for the homogeneous dynamics of the order parameter.

At this stage, one  can compute the steady state of $\psi$ by injecting Eq.~\eqref{eq:g:vonmises} into the r.h.s. of Eq.~\eqref{eq:dpsidt_ansatz_tot}, and equating the latter to zero.  One trivial solution is the isotropic state, of which we now discuss the stability by expanding, up to order~$\psi^3$
\begin{equation}
  \label{eq:dpsidt}
  \frac1\lambda \frac{\D \psi}{\D t} \simeq \bigl(\mu - D/\lambda\bigr)\psi - \xi \psi^3,
\end{equation}
with
\begin{align}
  \mu &:= \bigl\langle \pdp \bigr\rangle_0 , \label{eq:mu} \\
  \xi &:= \bigl\langle (\tfrac12 - \cos\Delta)\: \pdp \bigr\rangle_0, \label{eq:xi} \\
  \label{eq:avg0}
  \langle f \rangle_0 &:= \frac{1}{4} \int_{-1}^1 \D b \int_0^\pi\D \Delta\: \Bigl|\sin\frac{\Delta}{2}\Bigr| f(b,\Delta).
\end{align}
Altogether, the above calculation confirms our intuition that $\pdp$ is truly the relevant quantity to evaluate alignment in a system of self-propelled particles with binary interaction, provided that the density is low enough to ensure complete relaxation between successive scattering events.

\subsection{Vicsek vs. self-propelled disks alignment}
We shall now apply the above method to three different model systems. The motivation is to investigate how the local alignment rules affect the global aligning properties. We shall thus consider Hard Discs with Vicsek aligning rules, with inelastic collision, and with the rules identified above for the walkers.  In the first one, which we call the Vicsek Hard Discs (VHD) model, the continuous time noiseless dynamics is that of hard discs moving ballistically at constant speed $v_0$ and interacting with the Vicsek rules when they collide; the scattering parameter is the collision noise amplitude $\sigma$. The second case is that of Self Propelled Inelastic Hard Discs (SPIHD); the scattering parameter is the restitution coefficient $e\in\left[0 - 1\right]$. Between collisions, the velocity $\bfv_i$ of particle $i$ relaxes to $\hat\bfv_i=\bfv_i/|\bfv_i|$ on a timescale $\tau$. Here it is very clear that the collision conserves momentum, while the scattering event does not. The third case is that of the Walking Hard Discs (WHD) described by Eqs.~(\ref{eq:dotbfr}-\ref{eq:dotbfn}); the scattering parameters are $\alpha$ and $\tau_v$.\\

\begin{figure}[t]
\begin{center}
\includegraphics[width=0.98\columnwidth,trim=0mm 0mm 0mm 0mm,clip]{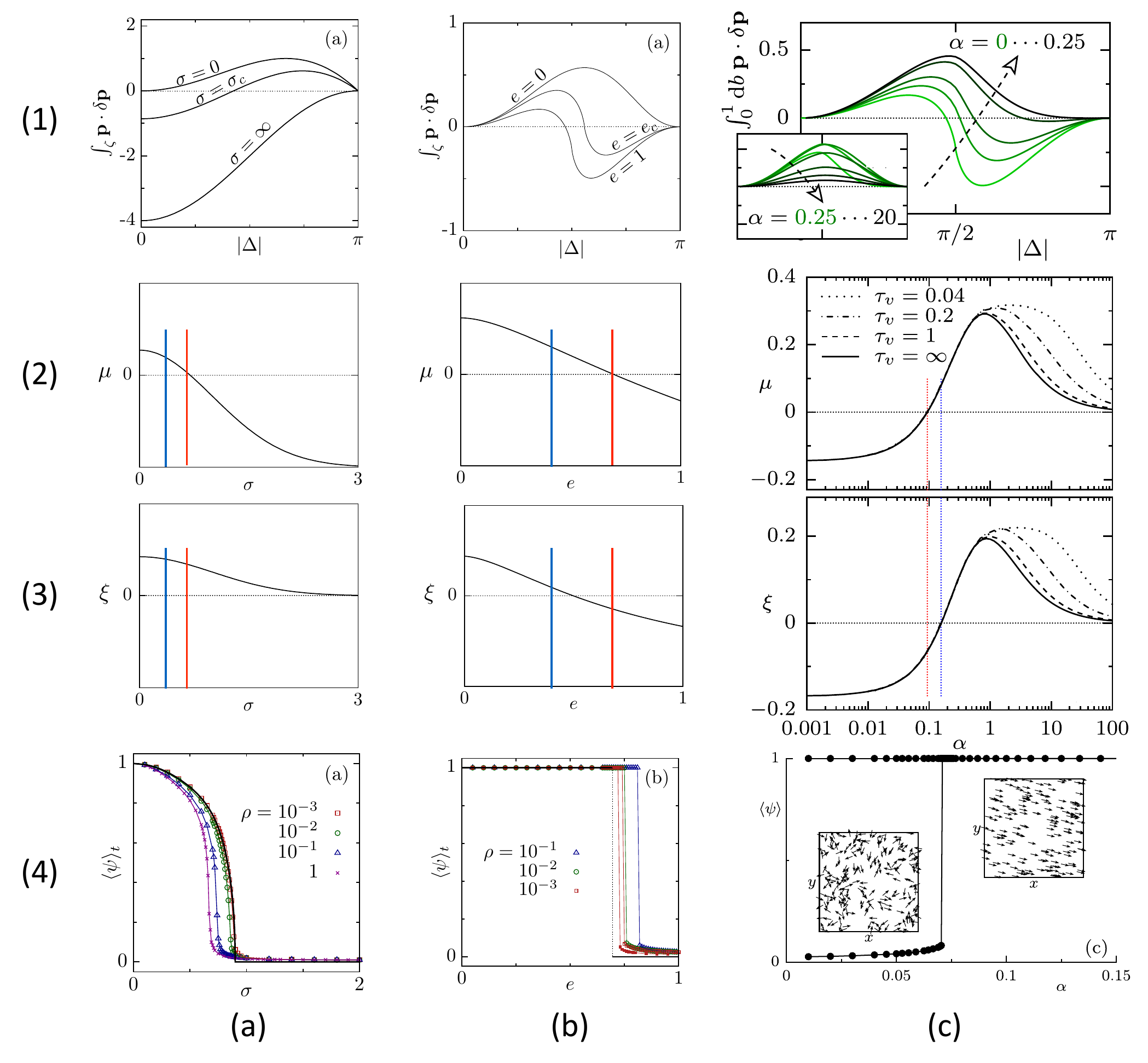} 
\end{center}
\caption{{\bf Transition to collective motion of self-propelled discs:} Columns (a),(b),(c) respectively correspond to the models VHD, SPIHD and WHD. Line (1) : the alignment $\int_\zeta\!\pdp$ as a function of the incoming separation angle $\Delta$ summarizes the microscopic dynamics averaged over the ``internal'' degrees of freedom of the scattering. Line (2) : the linear coefficient $\mu$ as a function of the internal scattering parameter; the red lines indicate where it cancels; the blue line indicate where $\mu - D/\lambda = 0$ for a finite value of $D$ and density $\rho=10^{-3}$. (3) : The third order coefficient $\xi$ as a function of the internal scattering parameter; the red and blue lines are reported from line (2). Line (4) : the order parameter $\left<\psi\right>_t$ as a function of the internal scattering parameter, in the absence of noise ($D=0$).}
\label{fig:pdp_align}
\end{figure}

The results are summarized in Fig~\ref{fig:pdp_align}, with columns (a),(b),(c) respectively describing the three models VHD, SPIHD, WHD. 
The first line provides information about the aligning property of the scattering event. 
The alignment function, $\int \D\zeta \, \pdp$, which depends only on the incoming angular separation $\Delta$, summarizes the microscopic dynamics averaged over the ``internal'' degrees of freedom of the scattering.
Consider first the VHD case, for which the ``internal'' degrees of freedom of the scattering is the collision noise. It is easy to see that $\pdp=|\bfp|(\cos\eta_1+\cos\eta_2-|\bfp|)$, where $|\bfp|=2\cos\tfrac\Delta2$. The integration over the collision noises is performed using $\int\D\zeta \equiv \int\D\eta_1 \D\eta_2 P(\eta_1) P(\eta_2)$, with $P(\eta)$ a gaussian
distribution of zero mean and variance $\sigma^2$. One obtains the alignment function
\begin{equation}
\label{eq:pdp_bdg}
    \int \D\zeta \, \pdp  = 2\cos\tfrac{\Delta}{2} \, \Bigl( 2e^{-\sigma^2/2} - 2\cos\tfrac{\Delta}{2} \Bigr).
\end{equation}
For $\sigma=0$ it is always positive, all collisions align on average; for $\sigma=\infty$ it is always negative, there is no alignment on average. At intermediate $\sigma$, collision with a large, respectively small, incoming angle separation $\Delta$ align, respectively dis-align.
For the SPIHD and WHD models, the alignment functions $\int_\zeta\!\pdp$ are computed numerically by simulating many binary scattering events at some fixed incoming angular separation $\Delta$, varying the impact parameter $b$ and are plotted on Fig.~(\ref{fig:pdp_align})-b1-c1. A central observation is that the alignment functions of these two models share important similarities, while they differ strongly from the VHD case. Indeed the Vicsek aligning rule is such that for large $\Delta$, it aligns the velocities as long as the noise remains finite; however for pairs of particles with velocities that are already well aligned (small $\Delta$), the noise dis-align them with high probability. Physical collisions between discs, and convex object in general, provide very different aligning rules. Consider for instance the case of the SPIHD displayed on Fig~(\ref{fig:pdp_align})-b1. Small incoming $\Delta$ favors alignment, and large $\Delta$ lead to dis-alignment. This is exactly the opposite physics.\\
We are now in position to discuss the impact of this important qualitative difference on the large scale physics, as far as homogeneous phases are concerned.  Computing the coefficients $\mu$ and $\xi$ now simply consists in averaging this function against the kinetic kernel $K(\Delta)$, including the geometric factor $\frac{1}{2}-\cos\Delta$, in the case of $\xi$.  The results are plotted on lines (2) and (3) of Fig.~(\ref{fig:pdp_align}). For the VHD case, $\mu$ is negative at large interaction noise $\sigma$ and turns positive for small enough $\sigma$. Similarly in the SPIHD case, $\mu$ is negative for large restitution coefficient $e$ and turns positive for sufficiently inelastic collisions (small $e$). The main difference is that in the VHD case $\xi$ is always positive and the transition is continuous as confirmed on Fig.~(\ref{fig:pdp_align})-a4. This is not the case in the SPIHD case. In particular $\xi<0$, at the transition where $\mu=\frac{D}{\lambda}=0$ in the absence of diffusion (red line). The transition is then discontinuous (Fig.~(\ref{fig:pdp_align})-a4). Interestingly, the presence of diffusion $D>0$ shifts the transition towards more inelastic systems, for which eventually $\xi$ becomes positive. There is therefore a first-order to second order transition driven by the amplitude of diffusion. In the WHD case, for $\alpha<1$, the situation is essentially identical to that of the SPIHD, the transition to collective motion being driven by the persistence of the polarity orientation $\bfn$. For $\alpha>1$, one observes (inset of Fig.~(\ref{fig:pdp_align})-c1) that the alignment function uniformly decreases towards zero: the persistence of $\bfn$ is such that collisions do not alter it anymore and alignment can not takes place. This explains the non monotonic $\mu$ and $\xi$ dependence on $\alpha$.\\

\subsection{Discussion}

Altogether the above analysis has revealed that the collective aligning strength within a system of self propelled particles is truly given by $\bigl\langle \pdp \bigr\rangle_0$. We shall however stress that the above derivation considers the situation where scattering events are well separated, so that the incoming particles have a common speed $v_0$, and fully relax their dynamics before a new collision takes place. This amounts to a low density approximation. Indeed one can see in line (4) of Fig.~(\ref{fig:pdp_align}), that increasing density from $\rho=10^{-3}$ to $10^{-1}$ already leads to quantitative shifts of the transition. These density effects are directly related to the renormalization of the coefficients $\mu$ and $\xi$ by the velocity fluctuations. Tackling this question is a significantly hard problem that has not been addressed yet.\\

We have also understood how purely dynamical alignment can take place and, in particular, how it can lead to collective motion in a population of self-propelled discs, with no steric source of alignment. On our way, we have seen that self-propelled hard discs, the alignment of which comes from physical collisions, behave very differently from their Vicsek counterpart in terms of alignment function, which in turn induces very different transition scenario between homogeneous phases. Let us stress here that this does not rule out the global Vicsek scenario, which at the end of the day, is dominated by nucleation like processes of the propagative bands. This is indeed what is also observed in the large scale simulations of the walking hard discs, as illustrated on Fig.~(\ref{fig:walkers})-e. One still expects that the differences observed in the alignment function may be significant when trying to compute the density effects we just mentioned. This in turn could play an important role on the precise nature of the localized non linear solution that are selected during the nucleation process. Here also is the frontier of the present knowledge in this matter.
 
\section{Perspectives}

In this chapter, we have focused on the transition to collective motion (TCM) in dilute systems of self-propelled particles, belonging to the class of polar and dry active matter. We have seen that for two very different systems, namely rolling colloids aligning through hydrodynamics interactions and walking discs aligning through dynamical relaxation of their polarity, the Vicsek scenario globally holds. It was however shown that such a conclusion can not be reached without a careful analysis of the relevance of the microscopic interactions. In the case of the rolling colloids, the long range dipolar hydrodynamics interactions could in principle suppress the giant density fluctuations predicted in the polar phase, although in the experiment they are too weak to fully destroy them. In the case of the walking grains, the structure of the hydrodynamics equation at the level of the Landau terms is very different from the case of the Vicsek model. It happens that the transition being governed by the coupling of the density fluctuations to the polar ordering in a nucleation like process, this last difference is not relevant for large system size. It remains however unclear wether it could become significant in the long range hydrodynamics properties, when considering the non-linear solutions.\\

Most theoretical results, discussed here, were obtain in the context of kinetic theory, and in the low density limit. Apart from the renormalization of the hydrodynamics coefficients by density and velocity fluctuations, there are good reasons to believe that density is a key control parameter in the physics of active matter, which does not only set the amplitude of noise above which collective motion sets in. 

First, we have mentioned in the introduction that, in the absence of alignment, a Motility Induced Phase Separation takes place in systems of self-propelled particles if the density induces a sufficiently strong slowing down of the typical velocity of the particles. This slowing down is generically induced by the crowding of the dynamics : the particles velocities decrease during their mutual interactions and the density is such that the relaxation towards the nominal velocity $v_0$ does not have time to take place. On one hand the same should hold in the presence of alignment. Then if the alignment is dynamically related to $v_0$, it could be enforced or suppressed.  Conversely, the alignment could prevent the particle from decreasing their velocity with density; in such a case the MIPS would not take place. Apart from a few papers~\shortcite{Farrell:2012vf,MartinGomez:2018wi}, the question of the coupling between TCM and MIPS remains largely unexplored.

Second, from the point of view of liquid state theory, how the activity alters the structural properties of a liquid and in turn its relaxation properties is essentially an open question, especially in the presence of alignment.
Finally, at large enough density, the active liquid is expected to eventually crystallize~\shortcite{Bialke:2012cw,Menzel:2013gs,Briand:2016fj}, or become glassy~\shortcite{Berthier:2013wg,Stuart:2013kt,Levis:2014ux}. The influence of activity on the glass transition has been investigated and it was shown that it depends on the stiffness of the interactions~\shortcite{Szamel:2015im,Berthier:2017te}. These studies were however conducted in the absence of alignment. How alignment contributes to this picture is far from simple. One could have thought that at very high density, the aligning processes would be destroyed by the high collision frequency. This is however not the case as exemplified by the observation of fully aligned active crystals flowing according to their boundary conditions~\shortcite{Briand:2018ew}.

\bibliographystyle{OUPnamed}
\bibliography{/Users/olivierdauchot/Documents/_Science/Biblio/Active}

\end{document}